\documentstyle[prl,multicol,aps]{revtex}

\renewcommand{\narrowtext}{\begin{multicols}{2} \global\columnwidth20.5pc}
\renewcommand{\widetext}{\end{multicols} \global\columnwidth42.5pc}

\begin{document}

\newcommand{\nt}{\narrowtext}
\newcommand{\wt}{\widetext}

\draft
\title{Indication of Superconductivity at 35 K in Graphite-Sulfur Composites}
\author{R. Ricardo da Silva\cite{R}, J. H. S. Torres, and Y. Kopelevich}
\address{Instituto de F\'{i}sica ``Gleb Wataghin'', Universidade Estadual de\\
Campinas, Unicamp 13083-970, Campinas, S\~{a}o Paulo, Brasil}
\maketitle

\begin{abstract}
We report magnetization measurements performed on graphite--sulfur
composites which demonstrate a clear superconducting behavior below the
critical temperature T$_{c0}$ = 35 K. The Meissner-Ochsenfeld effect,
screening supercurrents, and magnetization hysteresis loops characteristic
of type-II superconductors were measured. The results indicate that the
superconductivity occurs in a small sample fraction, possibly related to the
sample surface.
\end{abstract}


\nt
A considerable scientific interest of graphite and graphite-based
superconducting compounds \cite{Hannay,Koike,Belash,Chaiken,Jishi} has been
renewed \cite{Belashchenko,An,Baskaran} by the discovery of
superconductivity at 39 K in MgB$_{2}$ \cite{Nagamatsu}, a material similar
to graphite both electronically and crystallographically. Besides, recent
experiments \cite{Kopel1,Kopel2,Kempa} suggested the occurrence of
superconducting correlations in highly oriented pyrolitic graphite 
samples. It has been proposed \cite{Gonzales} that a topological disorder in
graphene sheets can trigger the superconducting instabilities.

In the present paper, we report an unambiguous evidence for the
superconductivity occurrence in graphite-sulfur (C-S) composite samples. A
clear superconducting behavior is found below the critical temperature T$%
_{c0}$ = 35 K.

The C-S composites were prepared by mixture of the graphite powder
consisting of $\sim $ 8 $\mu $m size particles [the impurity content in ppm:
Fe (32), Mo ($<$ 1 ), Cr (1.1), Cu (1.5)] and the sulfur powder (99.998 \%;
Aldrich Chemical Company, Inc.) in a mass ratio m$_{C}$:m$_{S}$ = 1:1 (m$%
_{C} $ = 0.5 g, m$_{S}$ = 0.5 g), where m$_{C}$ and m$_{S}$ are the
graphite and sulfur masses, respectively. The mixture was pressed into
pellets, held under Ar atmosphere at 650 K for 1 h and, subsequently,
annealed at 400 K for 10 h before cooling to room temperature. During
the heat treatment, 0.35 g of the sulfur was evaporated, i. e., the final
sulfur contents in the composite was 23 wt \%.

dc magnetization and low-frequency ($\nu $ = 1 Hz) standard four-probe
resistance measurements were performed on the sample of size 4.86 $\times $
4.52 $\times $ 3.52 mm$^{3}$ by means of SQUID magnetometer MPMS5 and PPMS
commercial equipment (Quantum Design).

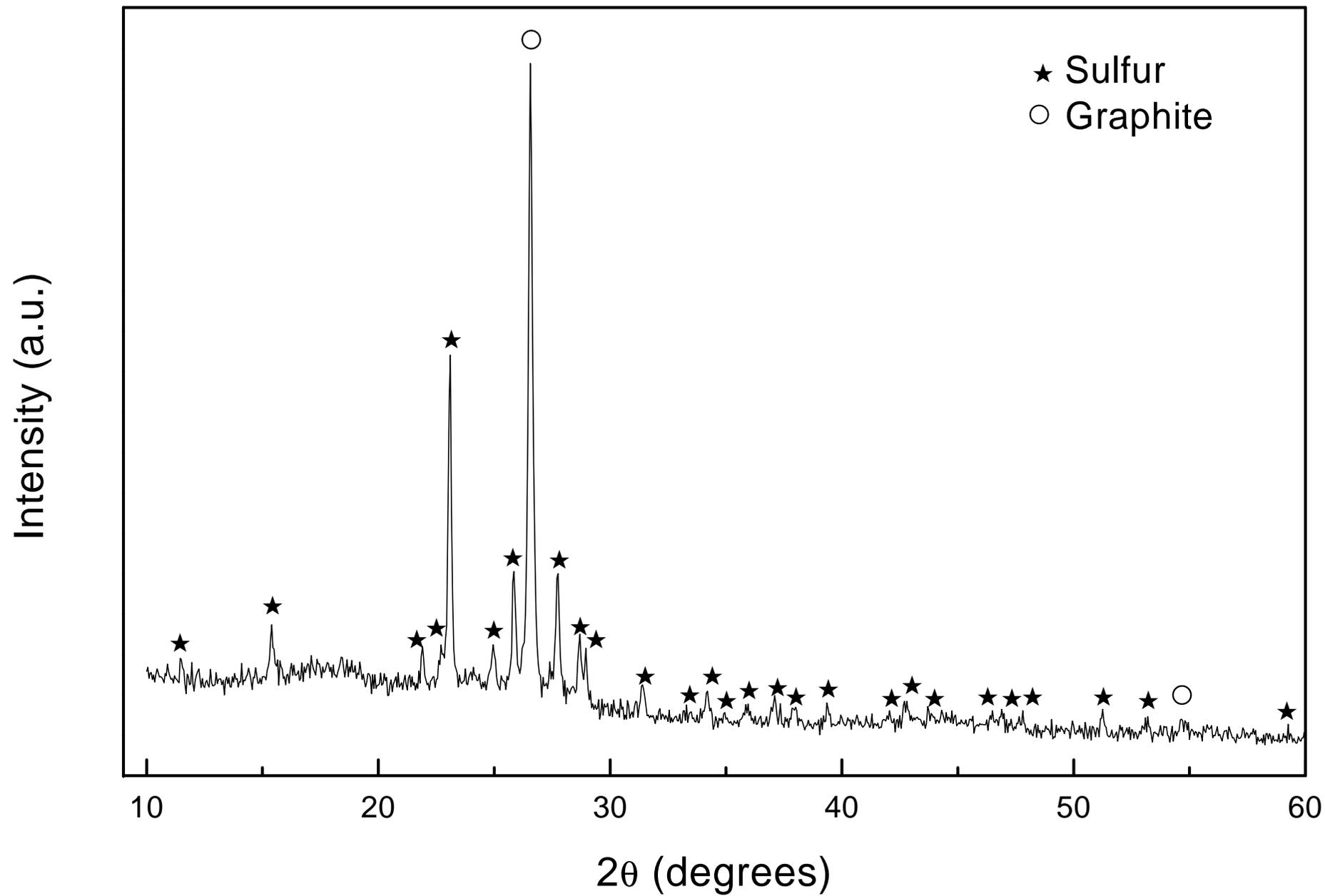
\begin{figure}
\caption{X-ray $\protect\theta $-2$\protect\theta $ diffraction pattern of
the C - 23 wt \% S composite.}
\label{lfig1}
\end{figure}

X-ray ($\theta $ - 2$\theta $ geometry) analysis revealed a small decrease
in the c-axis lattice parameter of the hexagonal graphite from c = 6.721 \AA
{} in the pristine graphite powder to c = 6.709 \AA {} in the composite
sample, and no changes in the lattice parameters of the orthorhombic sulfur
(a = 10.45 \AA {}, b = 12.84 \AA {}, c = 24.46 \AA {}). Figure \ref{lfig1}
shows x-ray diffraction pattern of C - 23 wt \% S composite obtained with Cu
K$\alpha $ source and 2$\theta $ step of 0.05$^{\circ }$.\ As Fig.\ \ref
{lfig1} illustrates, no impurity or additional phases were found.

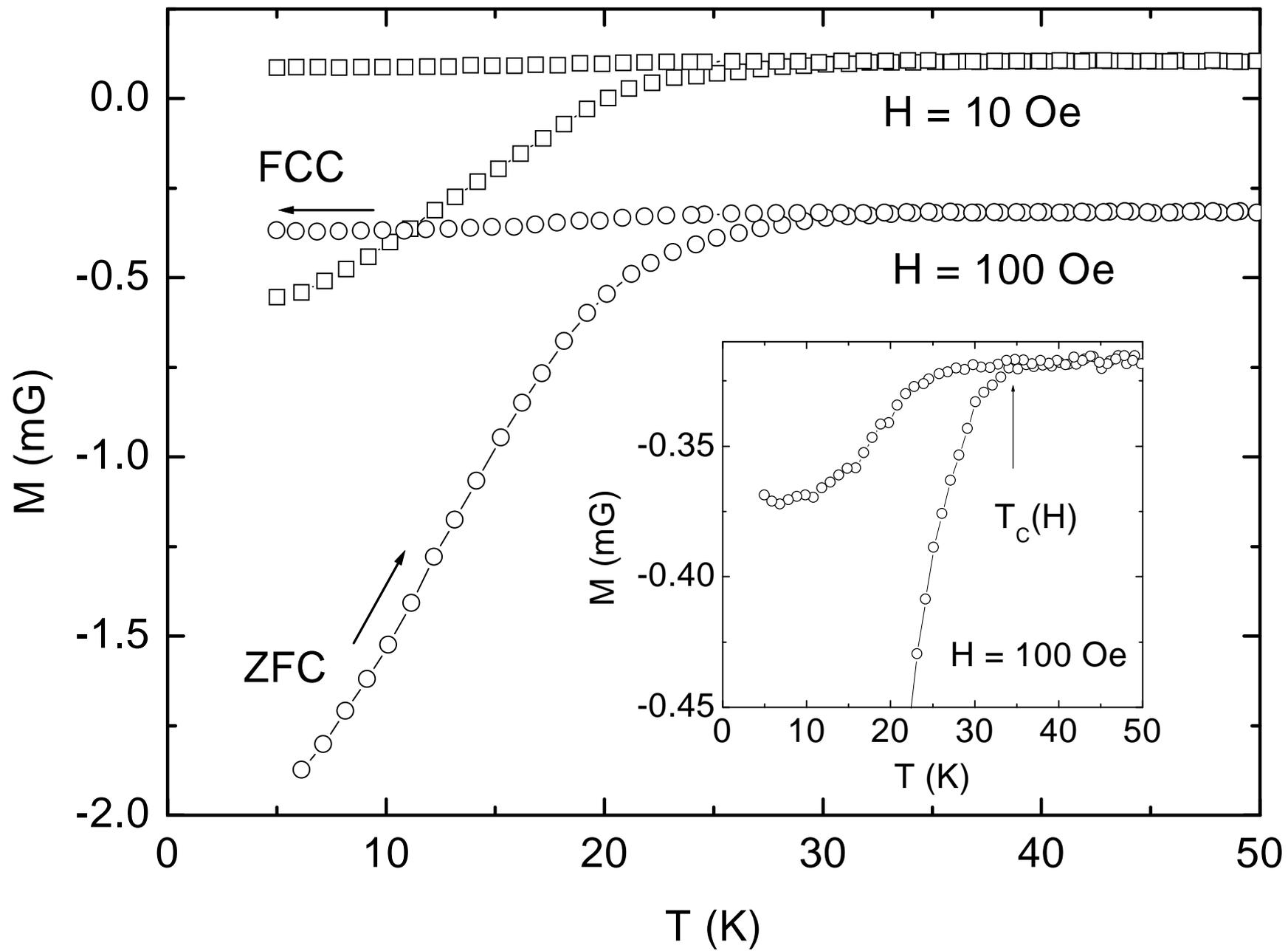
\begin{figure}
\caption{Temperature dependencies of the magnetization M(T) measured in
as-received sample A in zero-field-cooled (ZFC) and field-cooled on
cooling (FCC) regimes at two applied fields; 10 and 100 Oe. Inset gives
enlarged view of the superconducting transition recorded at H = 100 Oe.}
\label{lfig2}
\end{figure}

Figure \ref{lfig2} presents temperature dependencies of the magnetization
M(T,H) = m(T,H)/V (m is the sample magnetic moment and V is the sample
volume) measured in as-received sample (labeled here as A) at applied fields
H = 10 and 100 Oe. The magnetization data corresponding to the
zero-field-cooled (ZFC) regime, M$_{ZFC}$(T), were taken on heating after
the sample cooling at zero applied field, and the magnetization in the
field-cooled on cooling (FCC) regime, M$_{FCC}$(T), was measured as a
function of decreasing temperature in the applied field. Figure \ref{lfig2}
demonstrates a pronounced difference between M$_{ZFC}$(T) and M$_{FCC}$(T)
which occurs with the temperature decreasing. The inset of Fig.\ \ref{lfig2}
gives a detailed view of the data obtained for H = 100 Oe in a vicinity of
the T$_{c} $(H = 100 Oe) = 33 K below which a departure of M$_{ZFC}$(T) from
M$_{FCC}$(T) takes place. As can be seen from this plot, both M$_{ZFC}$(T)
and M$_{FCC}$(T) become more diamagnetic at T $<$ T$_{c}$(H). Such
magnetization behavior is characteristic of superconductors: The enhancement
of the diamagnetism below the superconducting transition temperature T$_{c}$%
(H) originates from the screening supercurrents (ZFC regime) and the
Meissner-Ochsenfeld effect of magnetic flux expulsion (FCC regime). It can
also be seen in Fig.\ \ref{lfig2} that, as the applied field increases, the
normal state orbital diamagnetism of graphite overcomes a positive
contribution to the magnetization (which can be due to both intrinsic weak
ferromagnetism of graphite \cite{Kopel2,Gonzales,Khveshchenko} and magnetic
impurities) resulting in a negative total magnetization above T$_{c}$.

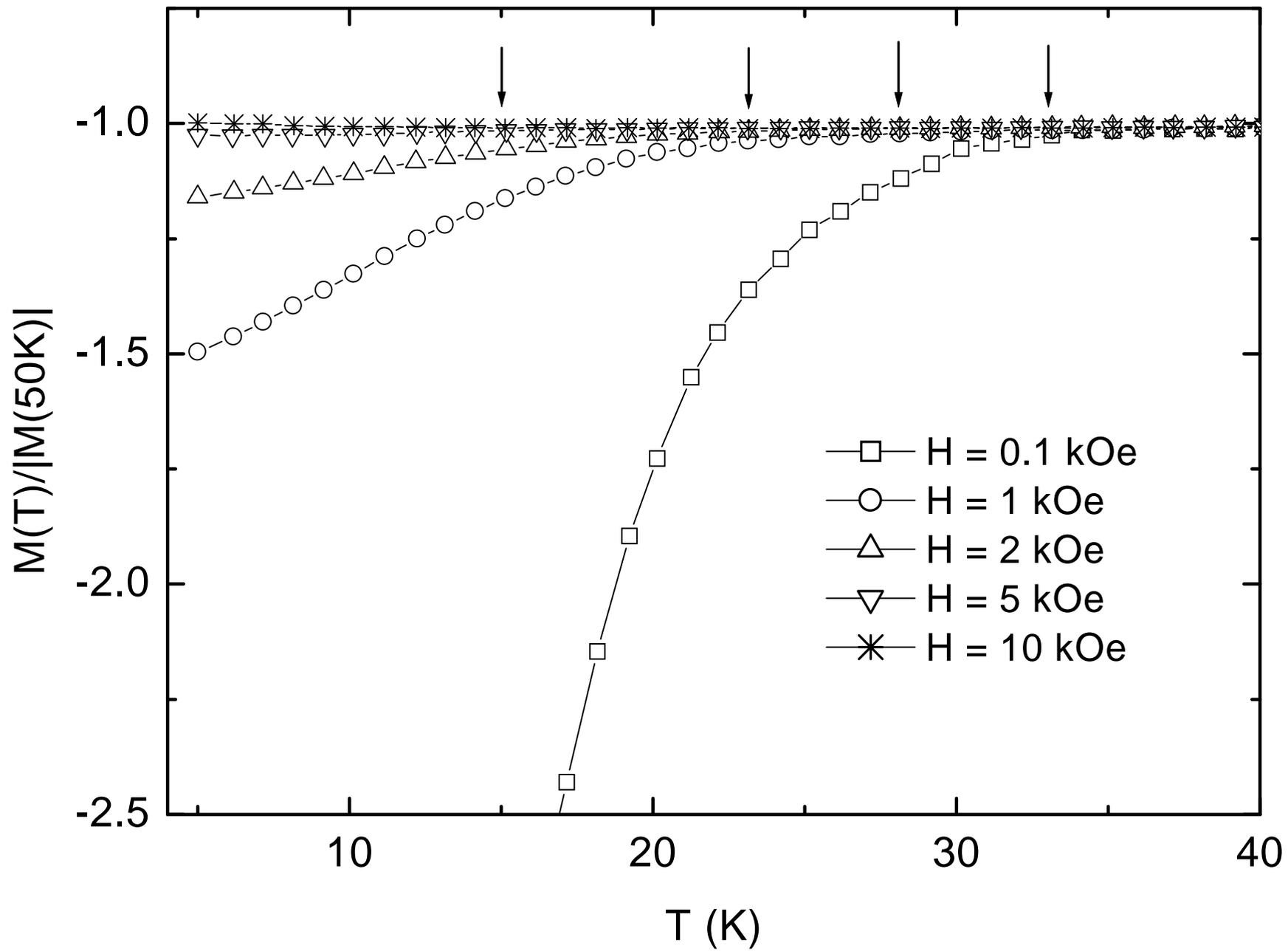
\begin{figure}
\caption{Normalized ZFC magnetization measured in sample A at various
applied fields. Arrows denote the superconducting transition temperature T$%
_{c}$(H) below which a departure of M$_{ZFC}$(T) from M$_{FCC}$(T) takes
place, as shown in the inset of Fig.\ \ref{lfig2}.}
\label{lfig3}
\end{figure}

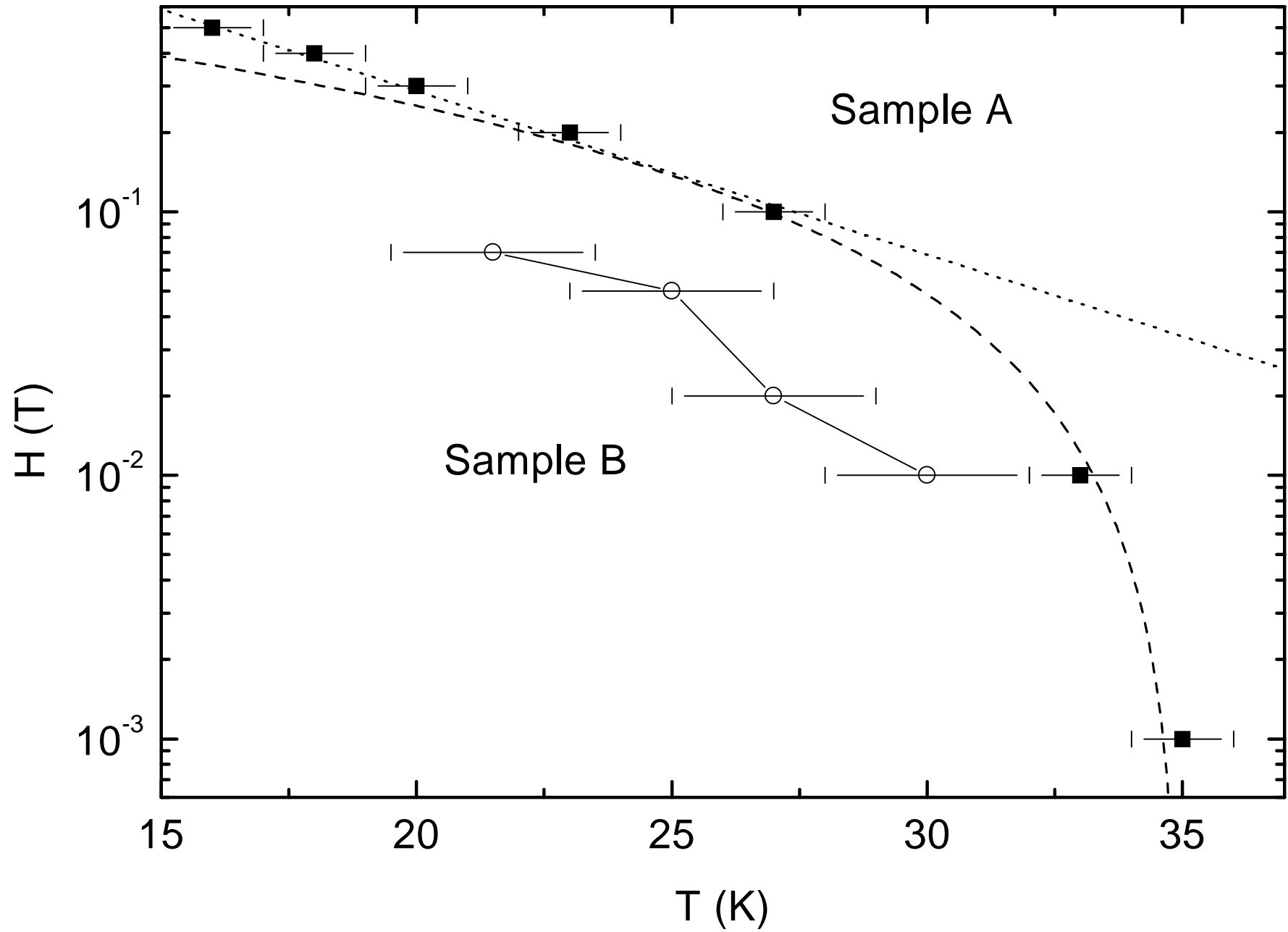
\begin{figure}
\caption{H(T$_{c}$) for samples A and B. Dashed and dotted lines are
obtained from Eqs. (\ref{eqn1}) and Eq. (\ref{eqn2}) with the fitting
parameters T$_{c0}$ = 35 K, H$^{\ast }$ = 0.9 T, T$_{0}$ = 7 K, and H$_{0}$
= 5 T.}
\label{lfig4}
\end{figure}

Figure \ref{lfig3} depicts the normalized ZFC magnetization M(T)/$|$M(50K)$|$
measured for various applied fields demonstrating that the transition
temperature T$_{c}$(H) decreases with the field increasing as well as that H
= 10 kOe completely suppresses the superconducting response. The obtained T$%
_{c}$(H) is given in the magnetic field -- temperature (H-T) plane (Fig.\ 
\ref{lfig4}).

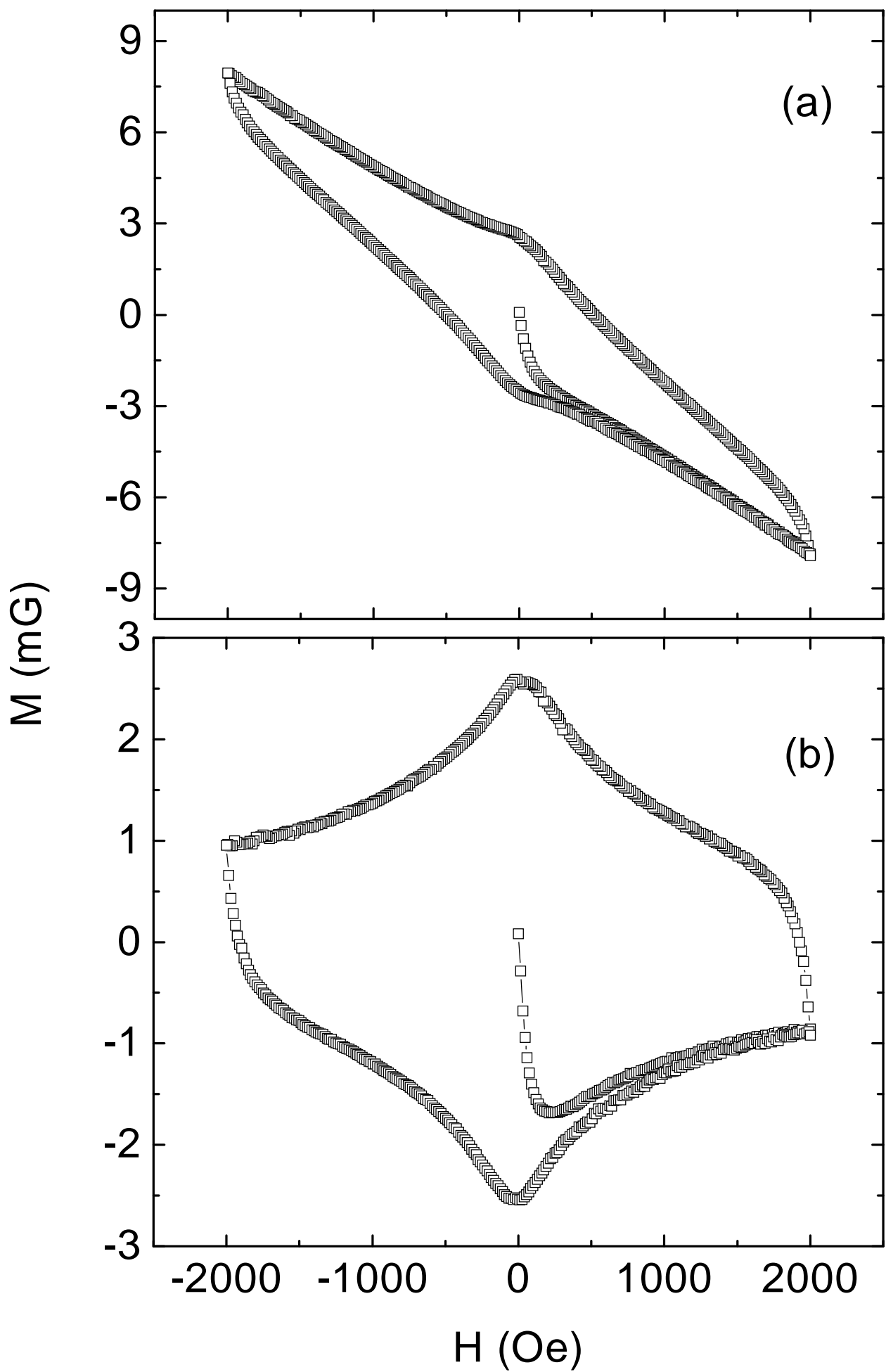
\begin{figure}
\caption{(a) Magnetization hystereis loop M(H) measured in the sample A at T
= 6 K; (b) M(H) obtained after subtraction of the diamagnetic background
signal M = - $\protect\chi $H with $\protect\chi $\ = 3.5 x 10$^{-3}$ mG/Oe.}
\label{lfig5}
\end{figure}

Figure \ref{lfig5} (a) presents the magnetization hysteresis loop M(H)
measured at T = 6 K after cooling the sample from 300 K to the target
temperature in a zero applied field. In Fig.\ \ref{lfig5} (b) we show the
same data after subtraction of a diamagnetic background signal. Figures \ref
{lfig5}(a) and \ref{lfig5}(b) provide unambiguous evidence that our sample is a type-II
superconductor with a strong vortex pinning \cite{Kim,Yeshurun}.

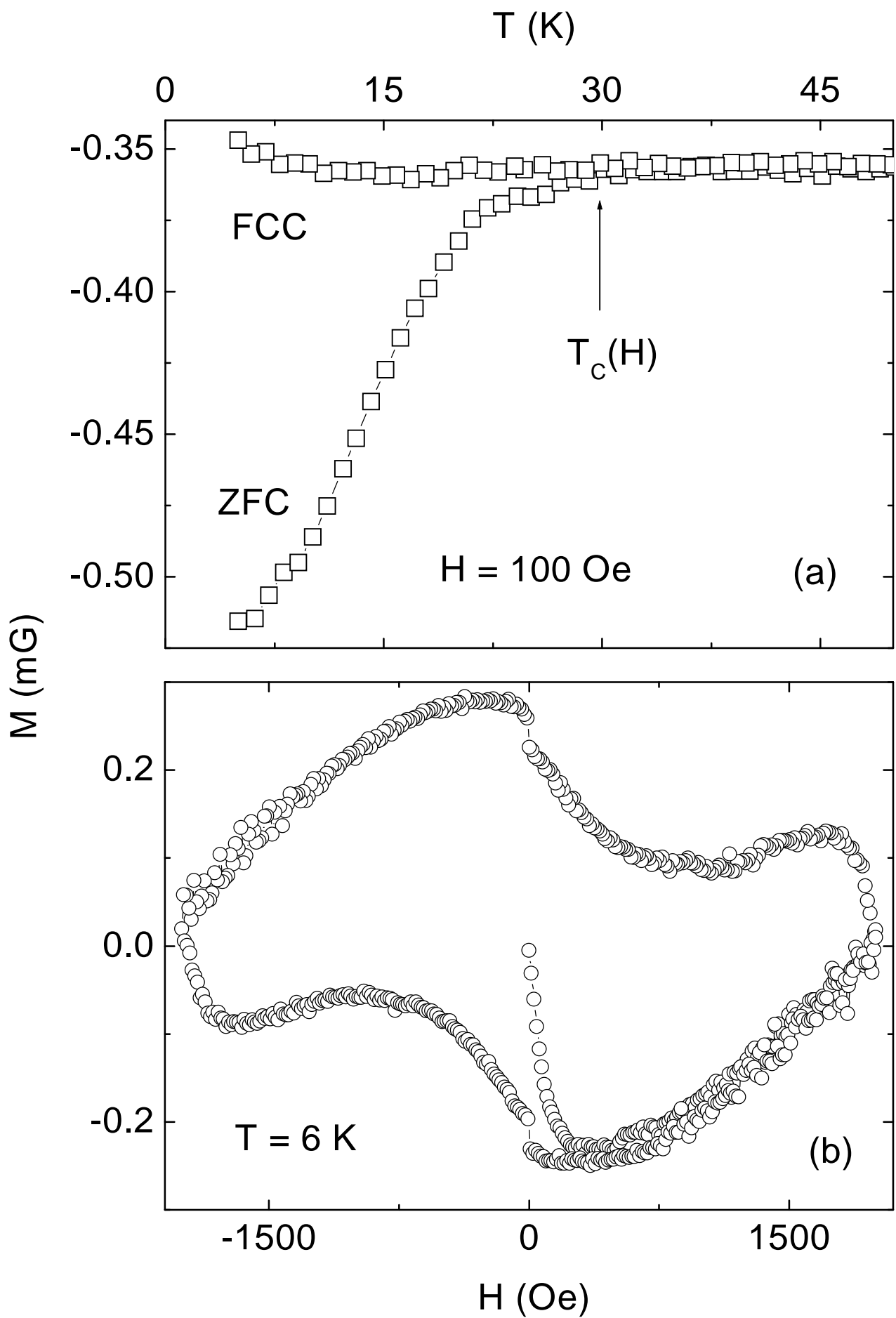
\begin{figure}
\caption{(a) M(T) measured in sample B in ZFC and FCC regimes at H = 100 Oe;
(b) Magnetization hysteresis loop obtained for the sample B at T = 6 K after
subtraction of the diamagnetic background signal M = - $\protect\chi $H with
$\protect\chi $\ = 3.5 x 10$^{-3}$ mG/Oe (6 K)}
\label{lfig6}
\end{figure}

In contrast to alkali-metal-doped graphite samples in which the
superconductivity vanishes after a short-time sample annealing at T $\geq \ $%
100 K \cite{Belash}, the superconducting properties in our sample were
stable during one week of measurements in the temperature range 5 K $\leq $\
T $\leq $ 300 K. To verify further the superconductivity stability, the
sample was kept at ambient conditions for two weeks. During this time the
sample has lost about 4 wt \% of sulfur. Then, we found a small decrease in T%
$_{c}$(H) and a strong reduction in magnitude of the superconducting
response. Figure \ref{lfig6} (a) exemplifies M$_{ZFC}$(T) and M$_{FCC}$(T)
recorded for this sample (labeled here as B) in applied field H = 100 Oe,
and Fig.\ \ref{lfig6} (b) presents M(H) hysteresis loop obtained at T = 6 K.
The transition temperature T$_{c}$(H) measured in the sample B for several H
is shown in Fig.\ \ref{lfig4}.

It is tempting to relate H(T$_{c}$) shown in Fig.\ \ref{lfig4}, to the upper
critical field boundary. However, a different interpretation is also
possible. It is found that H(T$_{c}$) obtained for the sample A can be best
described by the power law,

\begin{equation}
\stackrel{}{H=H^{\ast }(1-T_{c}/T_{c0})^{3/2},}  \label{eqn1}
\end{equation}
\mathstrut \bigskip in a vicinity of T$_{c0}$ = 35 K, and by the equation,

\begin{equation}
\stackrel{}{H=H_{0}\exp (-T_{c}/T_{0}),}  \label{eqn2}
\end{equation}
\bigskip \mathstrut 
below a reduced temperature T/T$_{c0}$ $\sim $ 0.8,
where H$_{0}$ = 5 T and T$_{0}$ = 7 K (see Fig.\ \ref{lfig4}). Equations (\ref
{eqn1}) and (\ref{eqn2}) imply that T$_{c}$(H) can be accounted for by the
existence of a breakdown field H$_{b}$(T) which destroys the
superconductivity induced by a proximity effect \cite
{Deutscher1,Deutscher2,Fauchere}. According to the theory \cite{Fauchere}, H$%
_{b}$(T) for normal-metal-superconductor structures saturates in the limit T 
$\rightarrow $ 0 to the value H$_{b}$(T = 0) $\approx $ 0.37H$_{0}$. Taking H%
$_{0}$ = 5 T, one gets H$_{b}$(T = 0) = 1.85 T which agrees with the
experimentally determined field H = 1 T at which the superconducting
response vanishes (see Fig.\ \ref{lfig3}).

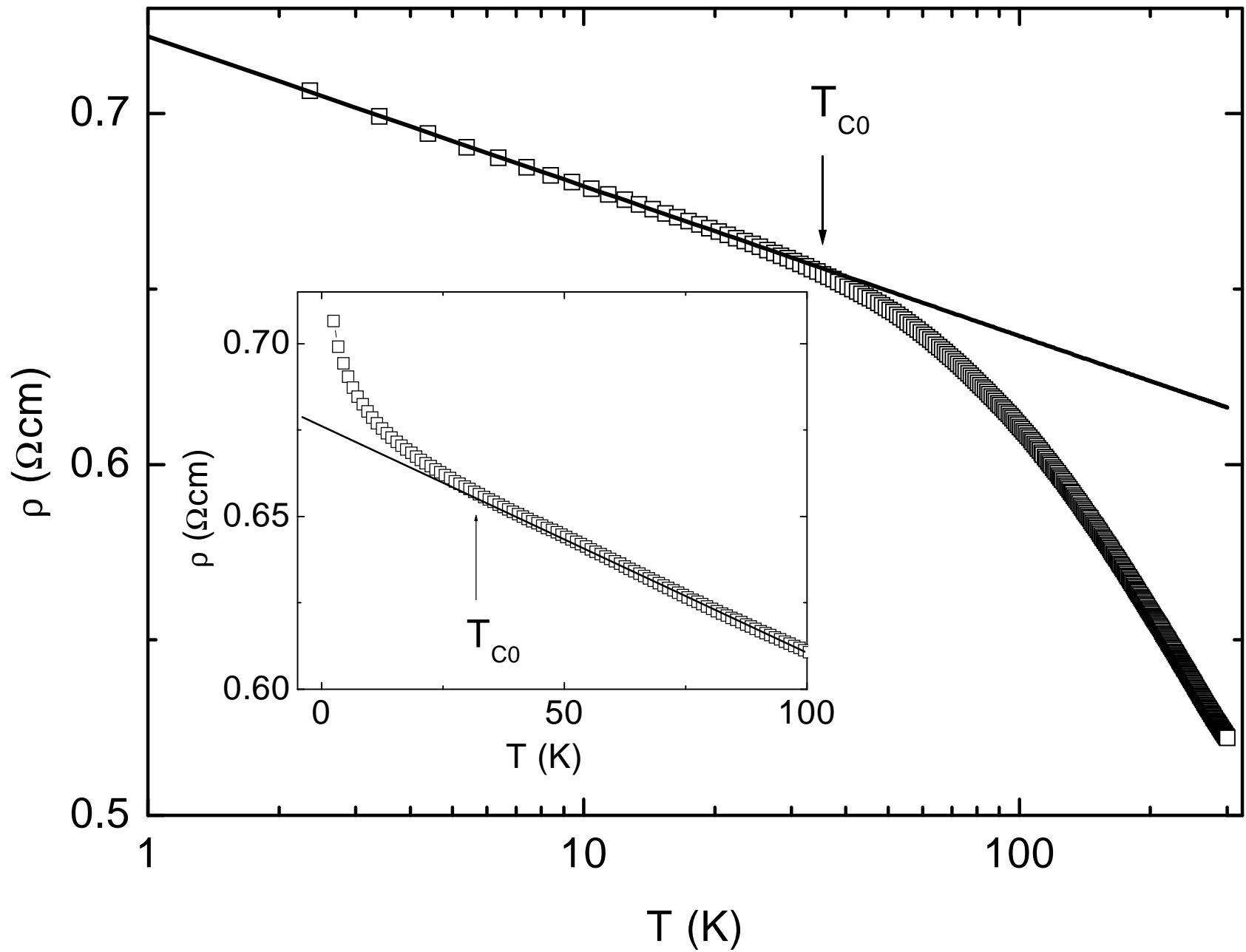
\begin{figure}
\caption{Semilogarithmic plot of zero-field resistivity $\protect\rho $(T)
measured in sample A; solid line is obtained from the equation $\protect\rho
$(T) = 0.6922 + 0.01855$\times $ln(T$_{0B}$/T), where T$_{0B}$ = 5 K. Inset
shows linear plot of the same data in a vicinity of T$_{c0}$ = 35 K; solid
line is only a guide.}
\label{lfig7}
\end{figure}

The zero-field resistivity $\rho $(T) measurements performed on sample
A (see Fig.\ \ref{lfig7}) revealed a slight increase ($\sim $ 20 \%) of the
resistivity lowering temperature from 300 K to T$_{on}$ $\approx $ T$_{c0}$
= 35 K, and its logarithmic divergence [$\rho $(T) $\sim $\ ln(T$_{0B}$/T)]
with a further temperature decrease, T $\leq $ T$_{on}$. The obtained $\rho $%
(T) at T $\leq $ T$_{c0}$ resembles the resistivity behavior in two
dimensional superconductors in the regime of weakly localized Cooper pairs 
\cite{Das}. While we cannot rule out completely other (fermionic)
mechanisms, the coincidence of T$_{on}$ and T$_{c0}$ as well as the obtained
low value of T$_{0B}$ = 5 K support the bosonic resistivity behavior \cite
{Das} at T $\leq $ T$_{c0}$.

Taking the magnetization and resistivity data together, one arrives at the
conclusion that the superconductivity in our samples is localized within
''grains'' or ''islands'' large enough to carry vortices. Note that the
magnetic hysteresis associated with the flux trapping can occur in materials
which consist of the superconducting elements each smaller than the
penetration depth \cite{Ebner}.

The magnetization data presented in Fig.\ \ref{lfig2} demonstrate that the
superconducting shielding effect is associated with a small volume fraction
of the sample ($\sim $ 0.05 \% of that expected for a bulk ideal
superconductor) which is compatible with the small size of the ''islands''.

We note further that T$_{c}$(H) measured in samples A and B differs by a few
Kelvin only (see Fig.\ \ref{lfig4}), whereas the superconducting shielding
effect as well as the magnetization hysteresis width are strongly suppressed
in the sample B. These results can be understood assuming that the size of
superconducting islands in sample B is much smaller than that in
sample A.

Finally, we stress that no sign of the superconductivity was found in our
pristine graphite powder. On the other hand, the highest T$_{c}$ = 17 K in
sulfur was reached under a pressure of 160 GPa \cite{Struzhkin}. We speculate
that the superconductivity in C-S composites originates from a sulfur-carbon
interaction at the graphite surface. Similar to the effect of adsorbed gases 
\cite{Lee,Ruffieux}, a hybridization between carbon and sulfur can increase
the local charge density and therefore trigger the superconductivity.
Further studies should verify this hypothesis.

In conclusion, the above results provide an unambiguous evidence for the
occurrence of high-temperature superconductivity in graphite-sulfur
composite samples and open new perspectives for the engineering of
graphite-based superconductors with high T$_{c}$. Recently, C-S composites
with T$_{c}$ exceeding 35 K have been obtained \cite{SergioM}.

This work was supported by FAPESP, CNPq, and CAPES Brazilian science
agencies.

\wt
\end{document}